# Evgeny P. Velikhov (1935 – 2024)

**Günther Rüdiger, Leibniz Institute for Astrophysics Potsdam**

The master student at Lomonosov University in Moscow E.P. Velikhov formulated 1959 the theory of magnetorotational instability, which dominates current astrophysics. A meteoric career made him later the science, nuclear and disarmament advisor to Gorbachev and Yeltsin. This article describes the interactions between Velikhov and the PROMISE team from Potsdam and Dresden-Rossendorf when it came to experimentally testing the theory in the laboratory in the 2000s. At an MHD conference in Catania, Sicily, he offered the public the use of small, transportable Russian nuclear power plants anywhere in the world until the fusion machines currently under development had finally solved the energy problem.

## 1. Potsdam (2004)

I was waiting in the arrival hall at Berlin-Tegel Airport for the Lufthansa flight from Munich, holding up the agreed signal, a large one-litre bottle of coca-cola, the kind I always used to keep in my car. A sprightly, round-headed passenger over 65 approached and introduced himself with a Russian accent: "Welichov". When I asked him about his short flight from Munich, he replied that he has "never flown economy before" and that there were a lot of people in his cabin. I considered this as a special kind of humour, took his bag and we walked to the car, which – what an airport – was less than 20 metres away. His remark, however, was meant seriously and was probably meant to imply that I had no idea who was standing in front of me. I had received news from Moscow that Evgeny Pavlovich Velikhov was staying at the Max Planck Institute in Munich and had contacted him there to bring him to Potsdam for a few days. I had sent him to Munich a standard second-class flight ticket and reserved a room at the usual three-star hotel in Potsdam-Babelsberg. I believed to have a needy Russian colleague in front of me who was delighted to receive an invitation to the newly western and generous Potsdam.

I had found the man who, as a graduate student in Moscow 45 years ago, had established a theory[1] that had dominated the current astronomical work for a decade and a half, and who had not even noticed this until I called him on the phone a few months ago. The astrophysical problem was to explain where the brightest cosmic candles, such as quasars, actually get the enormous energy that they radiate back as light. This gave rise to the idea of accretion discs around supermassive objects from which a lot of mass flows inwards if their viscosity is high enough to get rid of the angular momentum. However, in order to generate the enormous temperatures required for radiation, the viscosity must be so high that it appears to be caused by turbulence. The dilemma was that the imagined accretion discs with their Keplerian profile of rotation were by no means turbulent. All attempts to destabilise them in the frame of hydrodynamics, including our own, had failed. In 1982, during a visit to Potsdam,[2] the former nuclear weapons specialist Yakov Zeldovich asked "Why are these discs so turbulent even though they rotate stably? Are nonaxial disturbances or sound waves behind them? Find the solution here in this observatory!" I never forgot this request; he was in the right place, but we probably had other things to do at the time. The answer had already been given, hidden in the thesis of one of his close colleagues in Moscow, but forgotten by everyone, even the author himself. A rather similar problem had already been formulated a few years earlier in the frame of magnetohydrodynamics[3], but as a result of a certain, seemingly reasonable approximation, the magnetic field always had a stabilising effect, i.e. suppressing turbulence, which was a less interesting result. If Zeldovich had shared Velikhov's result with our uninformed group in Potsdam, many things might have turned out differently. Moreover, Zeldovich's name does not appear once among the now hundreds of citations of Velikhov's publication.

---

[1] E.P. Velikhov: J. Exptl. Theoret. Phys. 36, 1398 (1959).

[2] The leading design engineer Andrej Sakharov was only allowed to travel abroad during the Gorbachev era after decisive intervention by Velikhov.

As a master student Velikhov – adviced by the only 10 years older Stanislav Braginsky – had solved the well-formulated problem of what happens when an ideal fluid rotating between two cylinders is penetrated by a magnetic field along the axis of rotation. He found that the flow becomes turbulent and unstable when the outer cylinder rotates more slowly than the inner cylinder and the magnetic field is not too strong.[4] The paper which immediately was translated into English received little attention, achieving a total of 10 citations in the first 30 years, but today it has more than 600 citations. Velikhov's very first publication became by far his most successful one. On a whim, I had called him in Moscow to tell him about the fate of this publication, but ended up in reception rooms. Nevertheless, as it turned out, the message had reached him.

In the early 2000s, we wanted to find out in Potsdam why this phenomenon, now known as magnetorotational instability

---

[3] S. Chandrasekhar: Proc. Roy. Soc. A216, 293 (1953).

[4] The similar constellation with an azimuthal field always appeared stable as long as no current flows through the liquid.

(MRI), had remained unknown sofar in experimental physics. The question was quickly answered. The conductive liquid metals available for magnetohydrodynamic experiments, such as sodium, gallium or mercury, behave hydrodynamically like water, but conduct electricity much less efficiently than plasma; their magnetic Prandtl number is less than or equal to $10^{-5}$. Theoretical calculations of such flows are often successful in the so-called inductionless approximation with a Prandtl number equal to zero. Velikhov's instability, however, proved to be the exception, as it does not follow from the simplified system of equations. For finite but very small magnetic Prandtl numbers the flows do indeed become unstable, as we found out, but only for extremely high rotation frequency of the cylinders. This is why it had not been discovered experimentally sofar and why we shall also be unable to design an experiment. In a first investigation of this kind[5] we arrived at unattainable Reynolds numbers of almost 100,000 for liquid sodium as the conducting material; the exact value calculated later was even ten times higher. Unsure about the disappointing result, I held back this publication for almost a year to observe further developments. It was finally published in 2001, in the same month as the paper by Hantao Ji and his group in Princeton, which came to a similar conclusion. At the beginning of 2003, together with Dima Shalybkov from St. Petersburg, we published the final numbers calculated with a new and improved code, after they had already been presented the previous year in the proceedings of a conference in Evanston, Illinois. With the aggressive liquid sodium as the conducting fluid, a double-walled cylinder with a diameter of 40 cm would have had to be rotated vibration-free at a frequency of 20 Hertz which, unfortunately, was completely unimaginable.

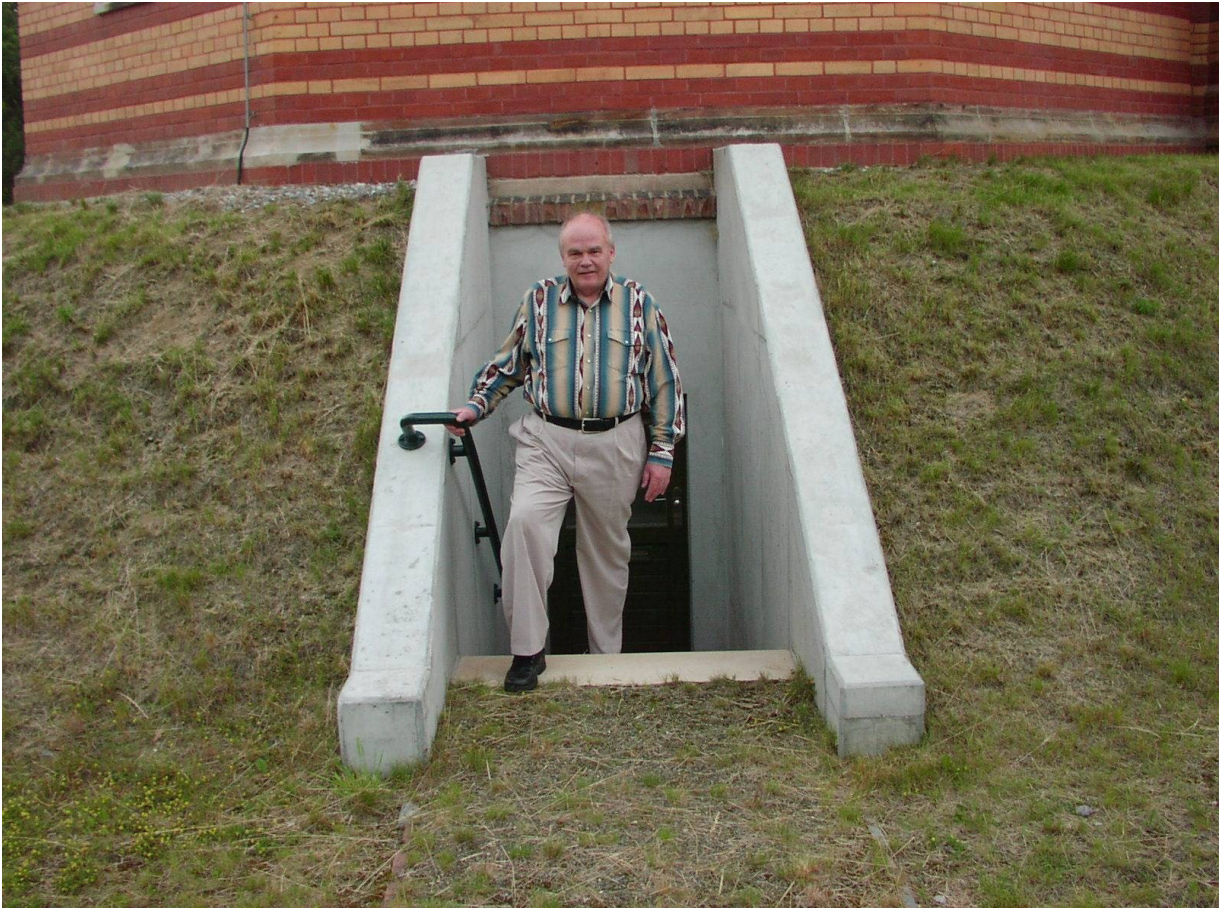

Figure 1 Evgeny P. Velikhov in Potsdam, Michelson-laboratory, Telegraphenberg

I called Rainer Hollerbach in Glasgow with whom I had published a common monography on "The Magnetic Universe". He was surprised to find that, numerically speaking, the "universe" now under consideration was only 40 cm in size. "Make it a little more complicated, with the field or flow spiralling like in a tokamak," he suggested. It can always be more complicated, I thought ironically, remembering the many additional epicycles that in earlier times astronomers had introduced in order to describe the planetary orbits with ideal circles, a misguided approach. After unsuccessfully investigating alternatives, the code was then converted to the more complicated magnetic background field, with stunning results. The young Velikhov had considered axial fields and azimuthal fields separately, but we had actually just thrown both together and treated real fluids between solid walls, which would not have been possible without computers. Suddenly also solutions in the inductionless approximation existed where only 1% of the original rotation frequency was needed. And, as a welcomed extra, the instability pattern moved along the axis, making it easy to observe. The numerical results even allowed the use of gallium instead of sodium as the fluid, which meant a huge cost reduction for a possible experiment. Part of MRI is that the liquid metal remains current-free on average; the magnetic field should not exert any force, it should only be there to act as a catalyst and it must not decay on its own. To achieve this, it was sufficient to pass a current within the inner cylinder, which created a ring-shaped vacuum field in the gallium that, together with the axial main field, will produce the needed spiral magnetic field.

On the one-hour ride to the Observatory Velikhov described his journey from Lomonossov University to the state-secret Kurchatov Institute and that he had previously visiting Max Steenbeck in Jena to discuss the construction of an MHD generator – whose villa was larger than his institute, and he described how difficult it had been to transfer the huge and formerly tolp-secret Kurchatov Institute into a public institution under the Russian Prime Minister after the end of the USSR. Yeltsin signed the far-reaching decree in a brief moment at Vnukovo Airport without reading it. He told me that he was in Munich for the international ITER[6] project and had taken two days off to learn in Potsdam what had become of his old work. He did not mention that it was he who had presented the paper to Yeltsin at the right moment and that he himself was the head of the Russian ITER delegation in Munich and the driving force behind the entire project. He did not mention this, but he did say that both his grandfathers had been arrested and shot during the Stalin era because they had been involved in illegal attempts to found a party or had tried to involve others in such attempts, now he could talk about this even abroad.

---

[6] ITER = International Thermonuclear Experimental Reactor basing on the Tokamak principle of Sacharov and Tamm.

After the usual round of introductions to the directors, who have little use for the guest's name and address, Velikhov wished to see the Michelson room in the lost Astrophysical Observatory. In fact, there is a replica of the instrument with which the young Michelson tried in vain to prove any anisotropy of the speed of light and – as this failed – was written off by his Berlin financiers. The instrument was so sensitive that it could not be installed in Berlin; only the strong pillars of the east dome of the observatory's main building on the sandy ground of the Telegraphenberg were suitable for its operation.

Velikhov's colloquium lecture was announced as "Astrophysics at the Kurchatov Institute" and presented the usual director's overview which is unlikely to leave a long-living impression. I introduced him as the then young inventor of the magnetorotational instability, which at first hardly anyone would have noticed, but which is now the focus of the work of many research groups, and mentioned that he comes from the famous Kurchatov Institute. Perhaps I felt uncomfortable introducing a purely astronomical contribution with too many career titles. His first sentence was that yes, he is from the Kurchatov Institute in Moscow, but it is also true that he is the president of this Science Center – probably the most powerful position ever held by a physicist in Russia. At this address, under the name "Lab 2", thousands of employees under the leadership of Kurchatov, Sakharov and Zeldovich had developed and refined the Soviet nuclear weapons. The heir to this gigantic military factory is now standing in front of German astronomers in the newly built Schwarzschild house of the Astrophysical Institute Potsdam – something that was impossible to imagine some years ago.

The next day was devoted to physics. The presentation of our results included our disappointment at the impossibility of realising standard MRI with the axial field, as well as our conviction that we could make progress with the spiral magnetic field. Velikhov thought in a different direction: there was no need to move the cylinders at all. If both are placed in a fixed position, an axially symmetrical current is generated from the outside to the inside, which, together with the axial main field, generates a Lorentz force that causes the liquid metal to rotate rapidly. However, it was argued that this so-called Dean flow has a disadvantage, because the rigid cylinder inevitably creates a super-rotation profile that increases from the inside to a certain maximum and which could significantly disturb the formation of the MRI. He asked whether we could probe this constellation, and we replied "Yes, we can".

Velikhov said goodbye at Tegel Airport telling that he would arrange for his new machine to be built according to our calculations the very next weeks. Could I promise to come to Moscow for its commissioning? In return, I promised to find ways to experimentally confirm his old discovery with our helicity concept. One of the two experiments would succeed, and then there might be some prize. With a sincere "Do swidania," I said goodbye and waved conspiratorially as he turned around once more on his way to security.

# 2. Dresden-Rossendorf (2005)

I knew Gunter Gerbeth and Frank Stefani from the Dresden-Rossendorf Research Centre – also a former nuclear research institute – from previous meetings. Immediately after Velikhov's visit, I asked them if they could imagine developing an experiment with sodium or gallium for a new variant of MRI. Of course, was the answer, but it's also a question of money. We had last met in early March 2004 in Catania, at a first conference on "MHD Couette Flows", convened together with Robert Rosner and Alfio Bonanno. The technical parameters were quickly put together. The outer cylinder was to have a diameter of 16 cm and a height of 40 cm. Inside the inner cylinder, a water-cooled copper rod was to run, capable of conducting currents of up to 10 kA. To calculate the influence of the two co-rotating lids, we needed a PhD student in Potsdam who redesigned a code for inductionless flows originating in England for our purposes. The total cost was around 500,000 EUR. In the summer of 2005, at a MHD conference in Riga, the concept for the machine, later named PROMISE, was announced, in which the cylinders would only have to rotate at frequencies of 1 Hz to excite the instability. After my presentation, E. Schartman from Princeton told me that all the inner cylinders in their containers had always remained solid, so that unfortunately no current could be passed through the axis just as it should be with us.

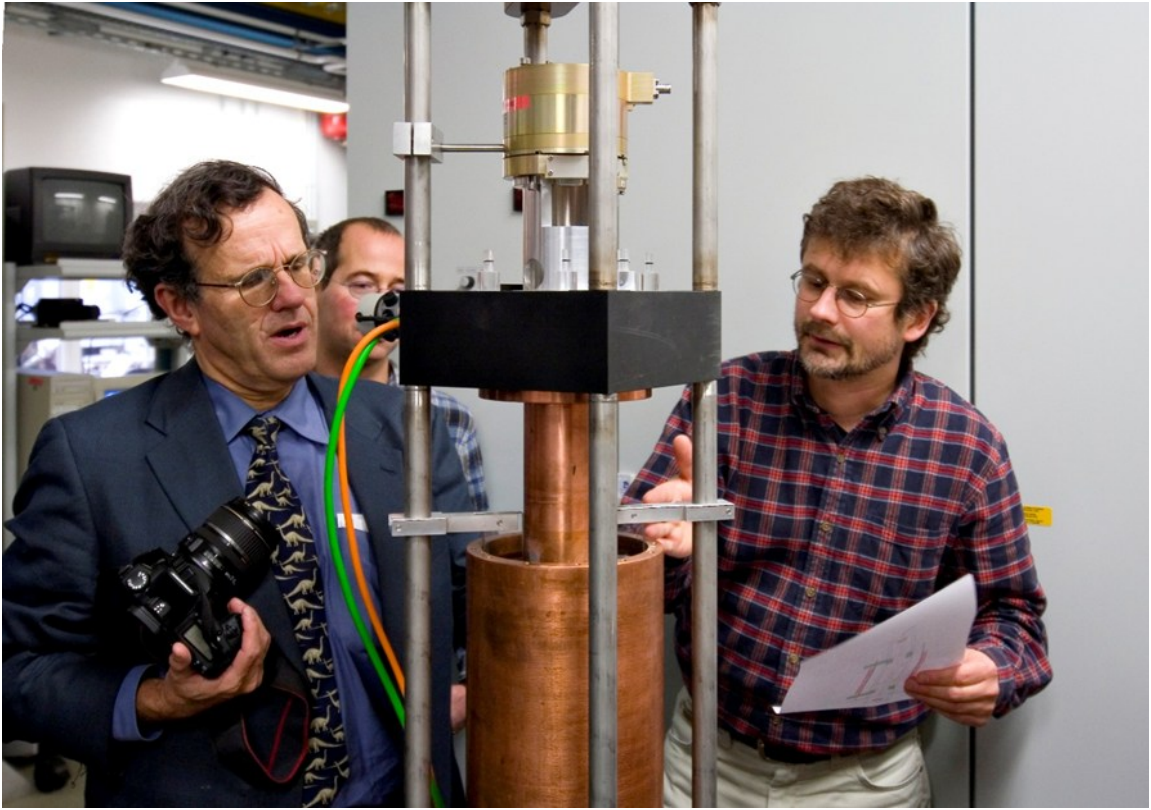

Figure 2 PROMISE im Forschungszentrum Dresden-Rossendorf. From left: R. Rosner, Th. Gundrum, F. Stefani

A brandnew science-project contest within the Leibniz Association, to which both the Potsdam Astrophysical Institute and the Dresden-Rossendorf Research Centre belonged, offered a source of funding. Each of the more than 80 institutes could submit one proposal per year for a novel and risky new project, whereby the funding requirements had to be considerable. Around 15 of these so-called SAW applications were approved each year, and our joint application, agreed in Riga, for a total of 400,000 EUR was approved on 24 November 2005 in an email from the Science Ministry. I had previously asked Robert Rosner from the Enrico Fermi Institute in Chicago to recommend PROMISE[7] as the AIP's competition entry for 2005 in a technical background letter to the institute's management. The disapproval I later received from the head, as well as the suggestion that we should rather build the MRI machine ourselves, had to be accepted with a shrug of the shoulders; it was indeed possible that alternative ideas had fallen behind because of our project.

---

[7] PROMISE= PotsdamROssendorfMagneticInStabilityExperiment

Intensive planning, rapid construction and swift experimentation took place in Rossendorf – there could not have been a better choice. The first results from PROMISE were published in journals such as Physics Review Letters and Astrophysical Journal[8] as early as 2006, partly because the axial wave propagation of the magnetic instability had made empirical verification much easier.

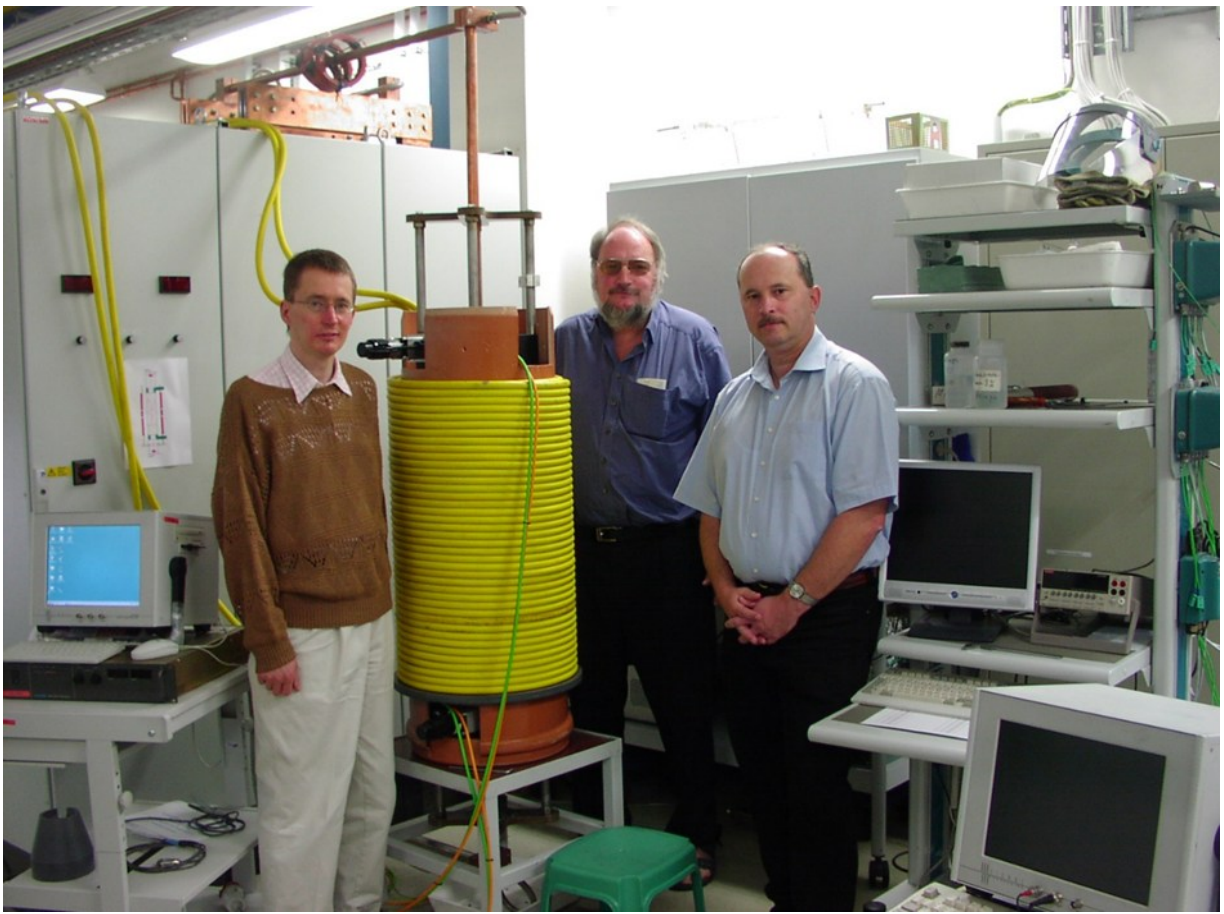

Figure 3    From left: R. Hollerbach, G. Rüdiger, G. Gerbeth

# 3. Moskau (2005/2006)

On Wednesday, 2 Feb 2005, I received a formal invitation to a festive reception on the occasion of Prof. Velikhov's 70th birthday on the coming Saturday in the ballroom of the Kurchatov Institute in Moscow. His assistant Andreev called me to say that his boss would greatly appreciate my attendance, that perhaps I could say a few words, that all expenses would be covered, that the visa would be issued on the same day when I arrived at the Russian embassy in Berlin – I only should have 400 EUR ready, which would also be reimbursed. During the night, I was plagued by clichés and old experiences with Sowjet organisations. What if no one picked me up at Sheremetyevo airport or, even worse, brought me back? What if the hotel was unsafe at night? What if the Kurchatov cash desk was closed over the weekend or just closed for no reason? On top of that, the only flight tickets available for Friday were on small planes from airlines I didn't know, costing more than €800. The next morning, they were already more than €1,000, and I didn't want to fly Aeroflot, which I had to suffer several times during the GDR era. If I hadn't cancelled the next day, the present report would certainly have been much more colourful. The celebration is said to have been iconic, but I have postponed my first visit to Velikhov by a short week.

On 10 February, I landed in the afternoon and I was picked up by a driver with a large cardboard sign and driven to Moscow in an old "Volga". At Khimki, at what used to be the entrance to the town, now seemingly in the middle of the city, the tank barriers from the WWII that once looked threatening were now just decoration, dominated by a huge IKEA department store. I was dropped off at a mighty building that served as a hotel or guest house for a former Soviet republic, with dinner and Russian breakfast, including sausages and cream. The hotel was a bit creepy at night because, as far as I could tell, I was the only guest. I was told I would be picked up at 10 a.m. the next morning.

In the morning, I was indeed picked up from the hotel on time with the engine running. I was driven through Moscow like a state guest in the back right-hand seat, then we went through an imposing and open gate, which was also used by pedestrians in both directions, past an oversized empty guard room with a conspicuous large label, and I was at the formerly top-secret Kurchatov Institute for Atomic Energy, now called the Russian Research Centre "Kurchatov". Velikhov had worked here continuously since 1961, after his time at Lomonosov University, becoming Director in 1988 and President of this institution during the Gorbachev and Yeltsin times. It was exciting to be driven through this mysterious gate without any checks. In his memoirs,[9] he recounts in detail that whenever there was a crisis, the authorities called on him for help, and so in 1992 he became President of both the Kurchatov Institute and the RosShelf corporation for the development of gas and oil reserves in Arctic Russia.

For unknown reasons I never saw Velikhov's office, perhaps because it contained a too large picture of the actual Russian President. I met him in a meeting room, together with his assistant A. F. Andreev and a younger scientist who was introduced as Vladimir P. Lakhin. I congratulated on the 70th birthday and presented him with our new monograph,[10] which included an extra chapter on hydrodynamic and magnetohydrodynamic instabilities in Taylor-Couette flows. He declined and instructed Andreev to pay for the book and the other expenses

[9] Evgeni P. Velikhov: "Strawberries from Chernobyl", www.createspace.com.

[10] G. Rüdiger & R. Hollerbach: „The Magnetic Universe", Wiley 2004.

immediately. Andreev took a wad of 50 EUR notes out of his trouser pocket and gave it to me; the amount was reasonable, but no more than that. A week earlier, I had been worried that the institute's cash register might be closed and I would not be able to be paid for my

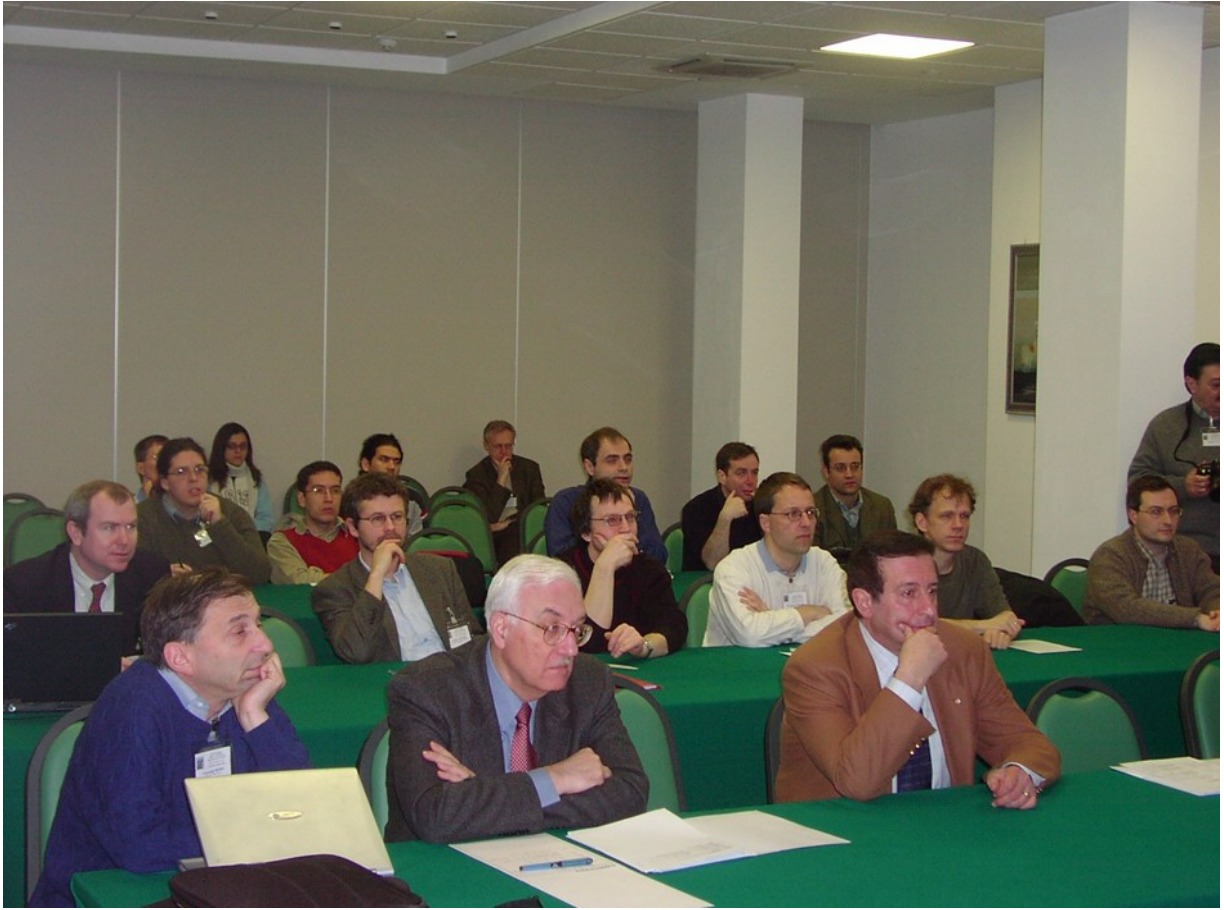

Figure 4  Lecture hall in Acitrezza (Catania) 2004, from left: Lathrop, Bodo, Hollerbach, Stefani, Ferrari, Cattaneo, Dobler, Bonanno, Belvedere, Brandenburg, Lanza (incomplete).

travel expenses as promised. Indeed, Russia was no longer Soviet Union.

My detailed powerpoint presentation was given to only five or six listeners and covered the full range of instabilities currently being discussed in Taylor-Couette flows, including magnetic instabilities and the recently intensively studied hydrodynamic stratotational instability (SRI), which occurs in density-stratified fluids under non-uniform rotation. The helical MRI underlying PROMISE was also briefly presented, but without mentioning technical details. The lecture led to two requests, both of which I accepted. Andreev wanted the presentation[11] for an anthology in honour of the jubilarian, and Velikhov suggested that Lakhin, who was present, may come to Potsdam to familiarise himself with the science. Vladimir did indeed come with own money for an extended period, and I also wrote and submitted the article devoted to Velikhov, but I never heard anything about the planned anthology.

One illustration from my presentation referred to last year's meeting north of Catania, for which Velikhov had sent a foreword to the proceedings.[12] We had invited him personally, but he did not come. I reported in detail on Hollerbach's calculations of the enormous influences of container lids, on Stefani's discussion of Taylor-Dean flows, and on the contributions of Lathrop from Maryland and Ji from Princeton. It was a description of a German-American competition that promised results in the near future. Velikhov must have been concerned, as he announced a workshop in Moscow for the following year, at which I was to give the opening speech. The official papers would arrive in good time he promised. If there were to be another meeting in Catania, he would certainly attend, as he had wonderful memories of Sicily, which he calls the cradle of civilisation in his memoirs, where there are more Greek temples than in Greece itself.

In the evening, there was a caviar banquet in the crowded ballroom at 1 Kurchatov Square. I sat on the presidium right next to Velikhov and said a few words by way of introduction, explaining that we were here because of an almost long-forgotten master's thesis that, unintentionally, contained a key to modern cosmogony, so wonderful was our science. It remained unclear to me who the many cheerful participants were, but if it was a mandatory event, then it was one with the best catering imaginable.

---

[11] G. Rüdiger: "Taylor-Couette flow: MRI, SHI and SRI", Astrophysics Data System 2005, unpublished (?).

[12] R. Rosner, G. Rüdiger & A. Bonanno: „MHD Couette flows: Experiments and models", Amercan Institute of Physics 733 (2004).

In October 2006, I indeed received a personal invitation to give a lecture at a scientific event at the Kurchatov Institute, without any information about the programme and/or other participants. I expected that Velikhov's own experiment would finally be presented, perhaps even with results, and prepared a lecture not only on the results with PROMISE, but also on the calculations for new experiments. In his old publication, Velikhov had not only investigated axial magnetic fields, but also considered azimuthal current-free fields, which in the simplest case always proved to be stable. However, we had discovered that there is also an MRI for azimuthal fields if more complicated non-axially symmetric disturbances are allowed. The existence of this initially numerically found "Azimuthal Magnetorotational Instability" (AMRI) was also demonstrated in a later experiment in Rossendorf.

In the same dark hotel as the previous year, I met Dan Lathrop for breakfast for two, but other activists from our industry did not show up. In fact, after we were taken to the lecture room at the institute, we realised that we were the only foreigners at the event. There were many younger people, some of them excited, other professors besides Andreev, a camera crew in the background, drinks, a short speech, and then I was told to begin. There was no time limit, and I don't remember a printed programme, but I had material on "Axial and Azimuthal MRI Experiments"[13] for just under an hour.

Dan Lathrop then explained his experiments at the University of Maryland; his presentation was much more technical than mine. Afterwards, a young man stepped forward and began by praising and thanking the academy members present, so I was sure I would hear news about the Moscow MRI experiment, but only brief theoretical contributions on instability theory with magnetic fields followed. At the beginning of his speech, the third of these speakers received a whispered instruction from Andreev, whereupon the introductory personal praise in the contributions disappeared from that point on. There was no information about Velikhov's experiment. Theoretical preliminary work[14] with the planned dimensions of the container (inner radius 3 cm, outer radius 15 cm, height 6 cm) had already been submitted as a preprint in April 2006. I had calculated the stability of this flow with stationary walls in Potsdam using our code without a lid and found that, due to the disruptive influence of the fixed inner wall, the magnetic field stabilises the flow for the small Prandtl numbers in question, i.e. suppresses MRI. The lids missing from the calculation represent the optimistic variant; their presence would only make matters worse, I argued, and sent the result to Moscow as agreed, but never received any response.

At afternoon tea time, Dan and I were driven along snow-covered side roads to Velikhov's private residence, a beautiful, simple villa in the middle of a high-rise neighbourhood, where the vehicles waited at a respectful distance. The house had been built by prisoners of war, we were told, and now looked very dignified. The host welcomed us in gardening clothes with his dog, and we were led straight to the study or library, where various chairs were arranged in a circle. Everything was rather cosy, and it remained unclear whether there were more representative rooms in the house.

---

[13] Rüdiger, Schultz, Szklarski (Potsdam), Hollerbach (Leeds), Gerbeth, Stefani, Gundrum (Rossendorf), Shalybkov (St. Petersburg).

[14] I.V. Khalzov, A.I. Smolyakov, V.I. Ilgisonis, E.P. Velikhov: "Magnetorotational Instability in Electrically Driven Flow of Liquid Metal: Spectral Analysis", *Physics of Fluids* 18, 124107 (2006).

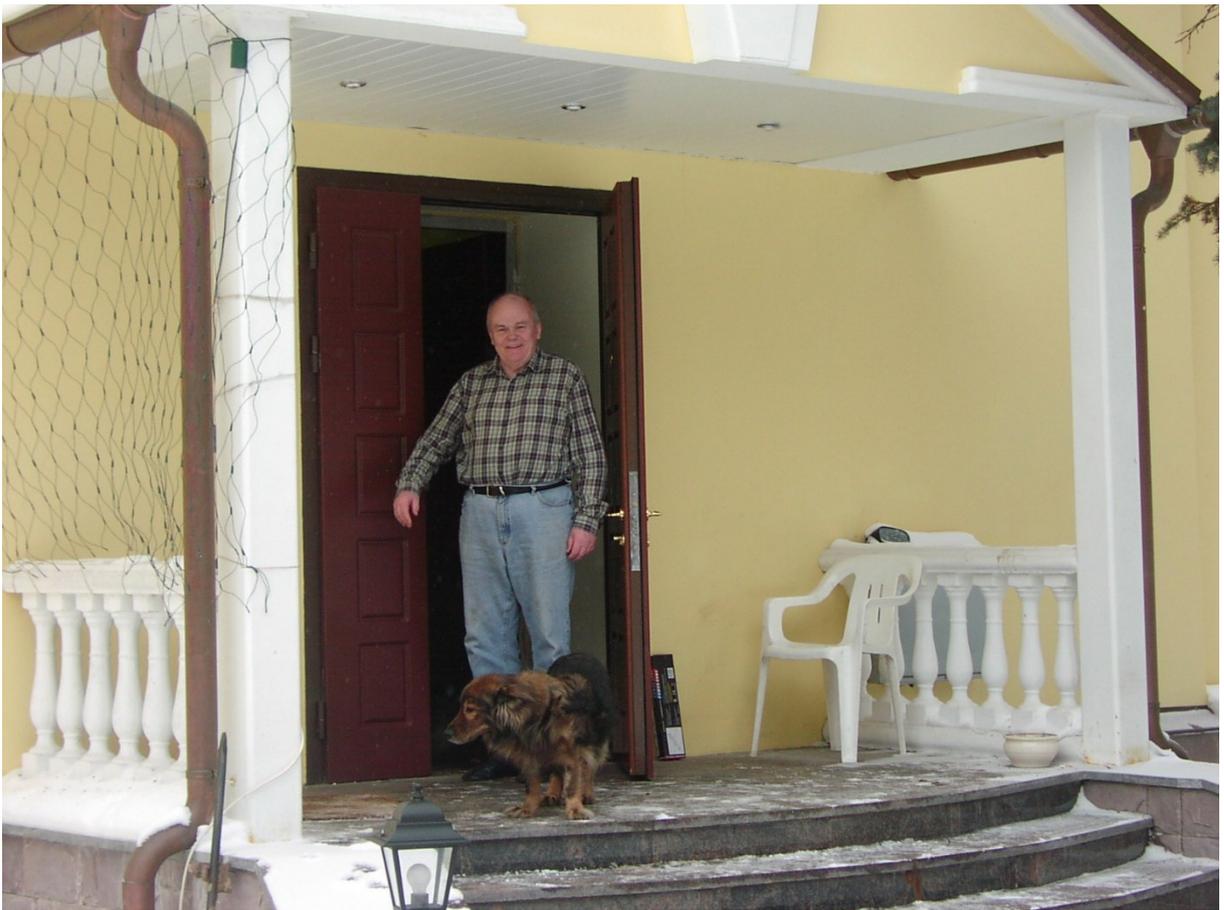

Figure 5 E.P. Velikhov with his dog at the door of his house in Moscow

Lathrop and Velikhov did most of the talking, mainly about measurement technology such as Hall sensors for magnetic fields and ultrasonic detectors for motion, to which I could contribute nothing. I tried in vain to bring newer experiments, such as those on Tayler instability, into the conversation, but in the end it came back to measuring devices, and it is possible that mutual agreements were being prepared. Regarding the Moscow MRI experiment, it was now said that there were technical problems at the design office outside Moscow, and it remained unclear whether the calculation results I had sent to Andreev had ever reached his boss. Velikhov's comment: It often happens in collectives that everyone is in favour of a project at the beginning, but later it is rejected. After tea and biscuits had been served, the elegantly dressed housewife appeared and we were introduced to her one by one.

I also visited a supermarket near the hotel. It was shocking, more than that, I was overwhelmed. Long shelves were crammed with products from all over the world, mostly packaged in the favourite colours of Russian customers, pink and light blue, strikingly similar to the colours of an American toy store. What a contrast to the Soviet shop with its puny offerings of bread, salt, tinned food and fish, presided over by a smoking cashier under a huge sign reading "Ne kurit".

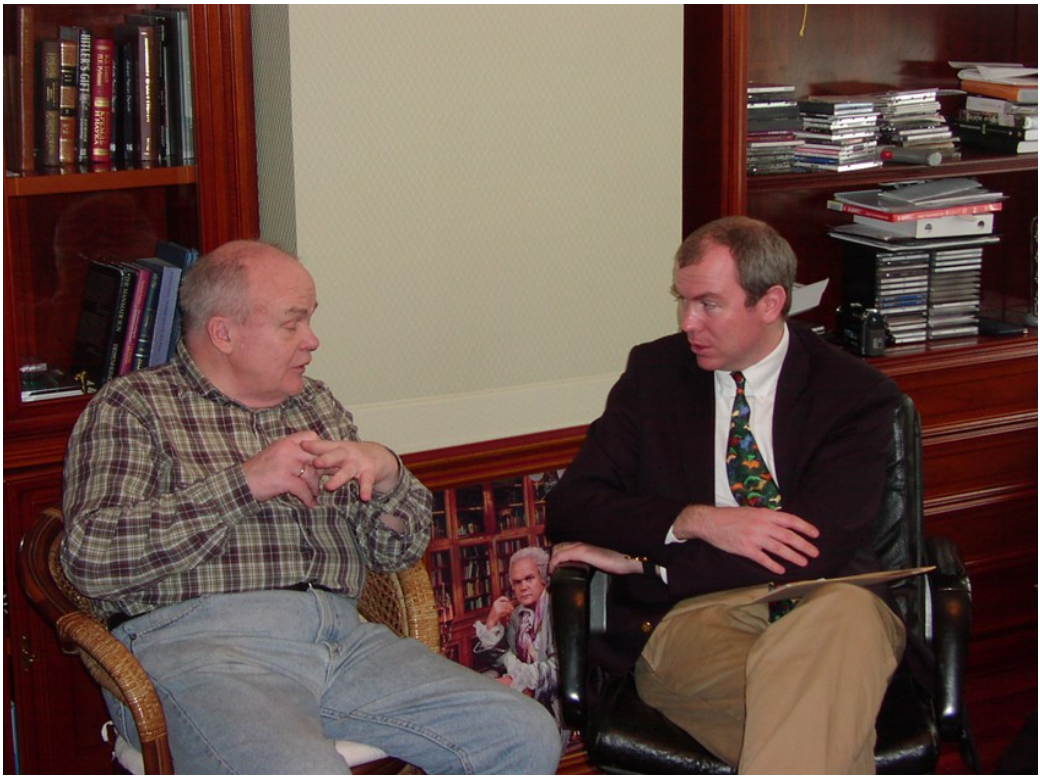

Figure 6   Evgeny Velikhov and Dan Lathrop in Velikhov's study

The closing dinner at the institute, this time with Velikhov surrounded by both the guests from the USA and Germany, was once again a highly refined affair, although half of the seats remained empty. As we now know, the Kurchatov Institute was facing a restructuring this year. It was during these weeks that the era of Putin's friend Kovalchuk as director and later president began, with new priorities that may have led to MHD experiments disappearing from the plans. At the same time, Velikhov, together with a Japanese and a French colleague, had just received the prestigious and highly endowed Global Energy Prize for 2006 from the Russian President "for the development of the scientific and technical basis of the ITER project". It was not until much later, in 2015, that Velikhov retired as honorary president of the Kurchatov Institute, which had been renamed the "National Research Centre". Over dinner, he told us that he had pushed through a doubling of the salaries of all Russian scientists in the Kremlin, which, after I later asked Leonid Kitchatinov from Irkutsk, had indeed happened.

# 4. Catania (2007)

At the beginning of October 2007, the Osservatorio Astrofisico di Catania hosted the second workshop on "MHD Laboratory Experiments for Geophysics and Astrophysics", this time in the city centre, at the Museo Diocesano in Piazza Duomo, overlooking the smoking Mount Etna along Via dell'Etna and located directly next to Catania's cathedral. In his welcome address, Alfio Bonanno listed the liquid metal experiments developed since Catania 2004 and announced newly constructed Couette flows. The aim of the workshop was to understand the obstacles facing current efforts to research magnetic instabilities in the laboratory. While in 2004 in Catania the question was still what an ideal MRI experiment would look like, with the one positive answer that the critical Reynolds number for the occurrence of MRI decreases dramatically when the purely axial field is replaced by a spiral field. The known experiments would cover an impressive range of applications. These often concerned the stability of differential rotation, the DTS experiment in Grenoble, the VKS dynamo experiment in Cadarache and the announced experiment in Moscow. On the other hand, dynamo experiments were running in Riga, Madison, Maryland and Perm, where it is now even possible to measure magnetic turbulence diffusivity. He mentioned a recent result of Peter Frick from Perm, where the effective electrical conductivity was determined in a turbulent flow of liquid gallium in a closed toroidal channel. The maximum deviation of the electrical conductivity from its Ohmic value was small but definite.

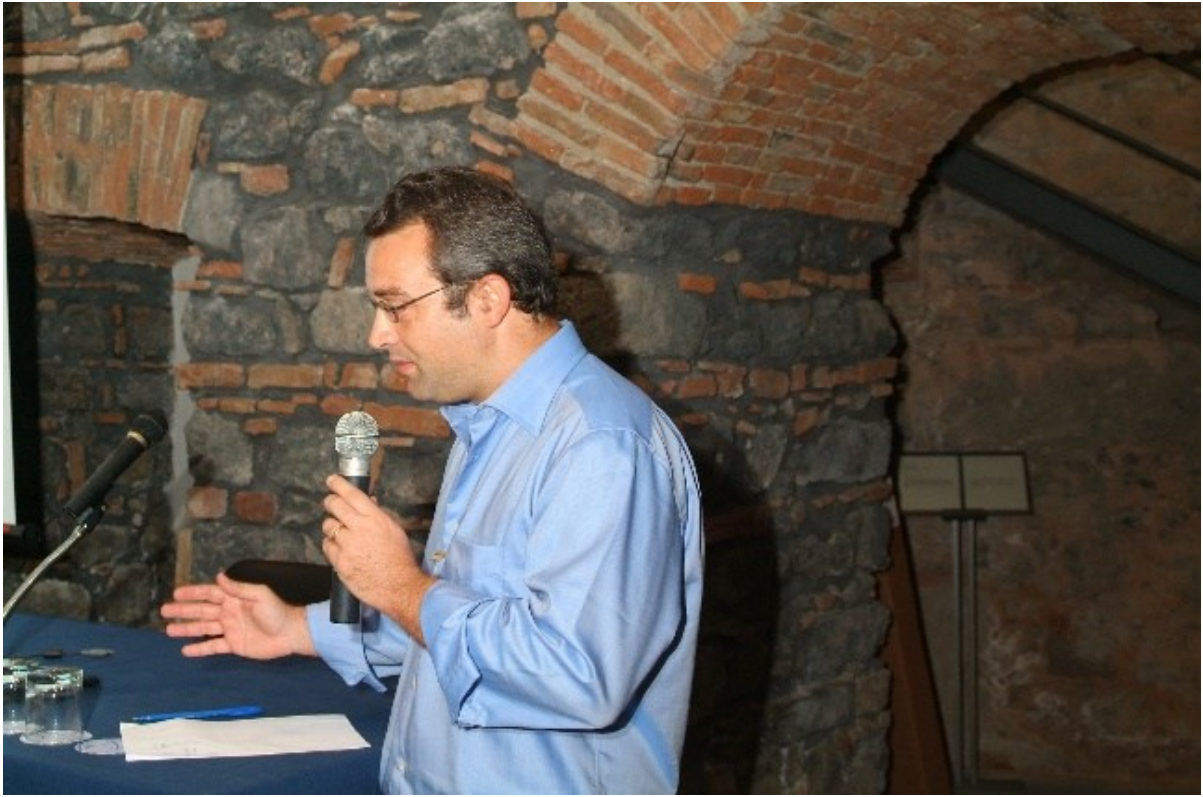

Figure 7 Alfio Bonanno opens the Catania Conference 2007

Experimenters from all active laboratories had travelled to the conference (Cottbus, Los Alamos, Lyon, Maryland, Meudon, Princeton, Riga, Rossendorf), as well as theorists from Beer Sheva, Berkeley, Chicago, Leeds, Potsdam and Stockholm. Velikhov had made good on his announcement and arrived in Catania from Brazil. He opened the MRI meeting of the conference with a presentation of an extensive collection of formulas. The results of Lakhin, who had carried out comprehensive calculations in shortwave approximation in Potsdam, were also presented. The presentation seemed impersonal and poorly prepared, and once again the Moscow experiment was hardly mentioned, if at all.

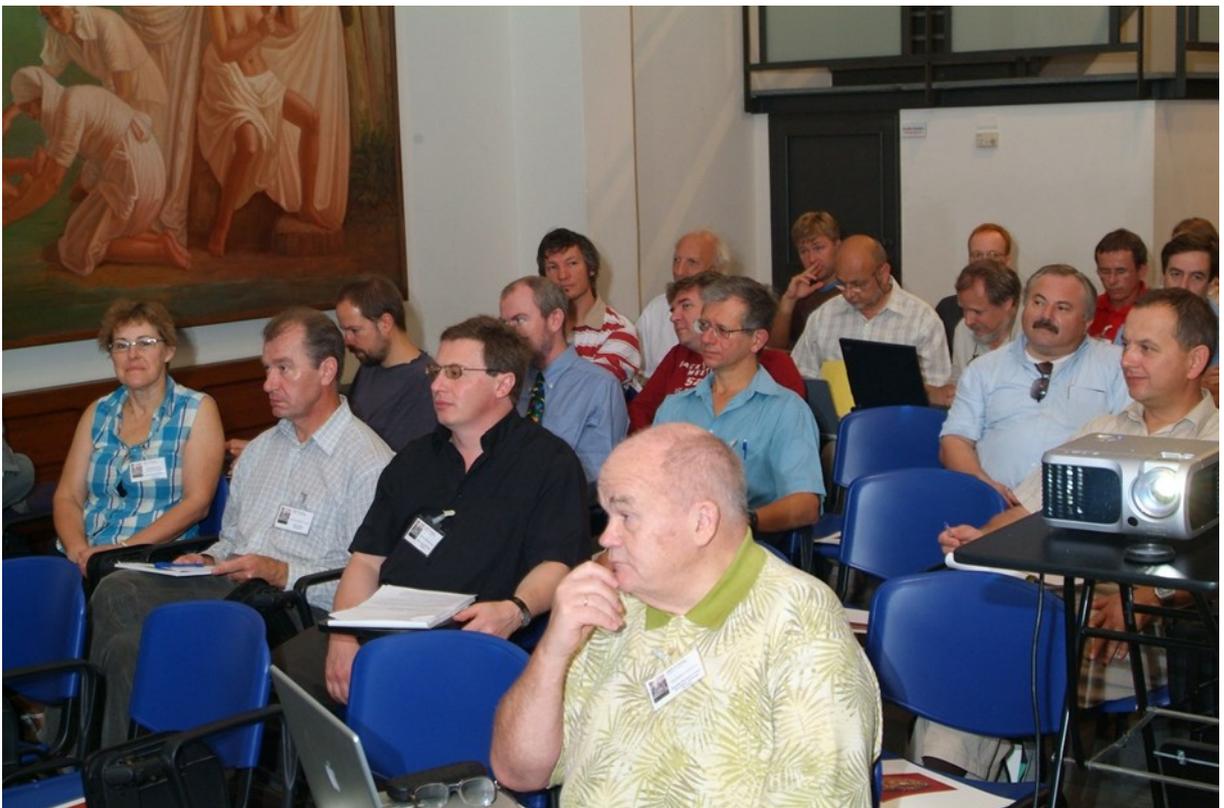

Figure 8 Catania 2007, from left: Otmianowska-Mazur, Kitchatinov, Giesecke, Rogachevskii, Lathrop, Velikhov, Gailitis, Nataf, Szklarski, Leorat, Gellert, Kleeorin, Elstner, Hanasz, Cattaneo (incomplete)

Stefani then summarised the design, measurement technology and results of the PROMISE experiments. The spiral-shaped MRI is observed as a travelling wave with the predicted phase velocity using ultrasonic sensors, tested on Taylor vortices, which always occur without a magnetic field when the outer cylinder is at rest. He presented two series of experiments with different electrical boundary conditions, one with conductive cylinders and one with insulated cylinders. As expected, the helical instability developed more easily with highly conductive cylinders. A radially concentrated flow appeared in the centre between the lids, in which the two Ekman vortices from above and below met. The insulated inner cylinder helped to suppress this phenomenon. In a later design ("PROMISE II"), the upper lid of the copper container was cut open in a ratio theoretically determined by our Polish PhD student Jacek Szklarski in order to further reduce the Ekman vortices.

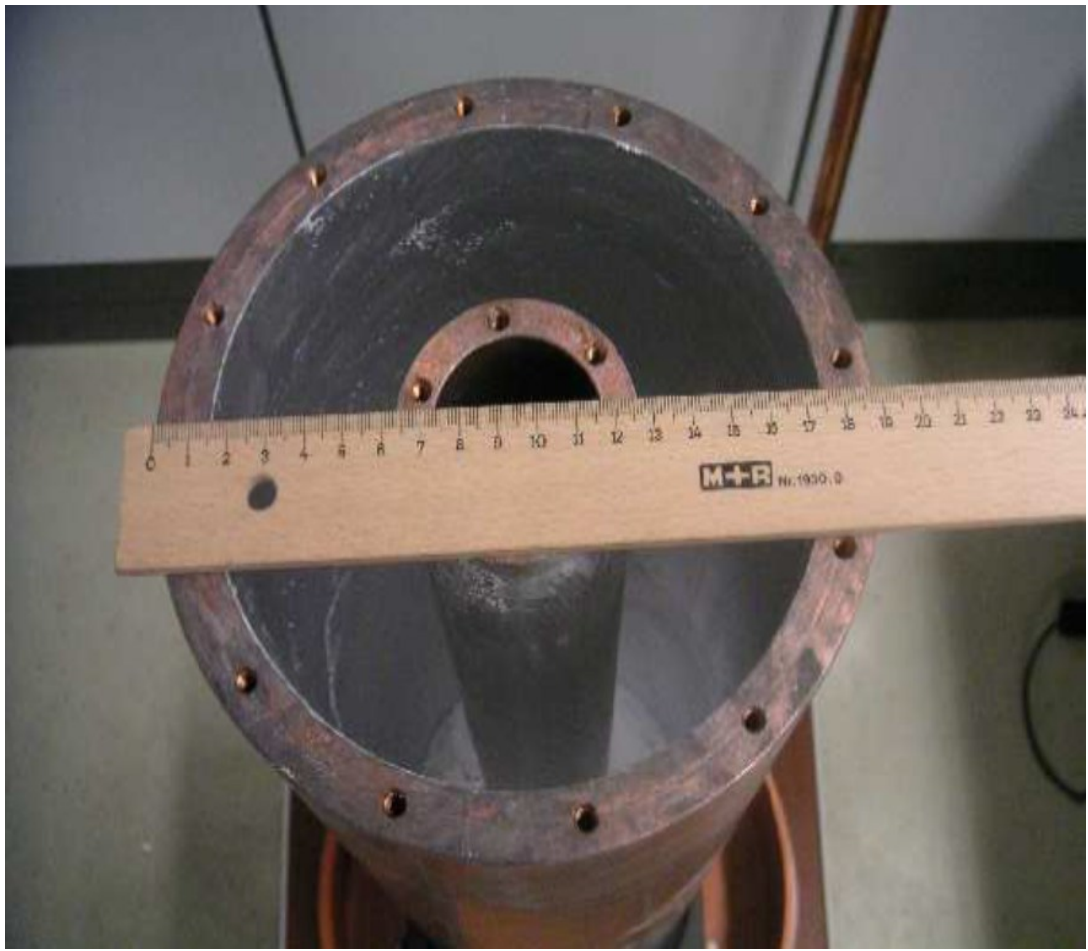

Figure 9 The PROMISE container from above, lecture by Stefani (2007)

Hantao Ji from the PPPL[15] addressed the recently raised hypothesis that even purely hydrodynamic instabilities could cause the necessary viscosity values in accretion discs, which would render the MHD experiments and simulations presented here astrophysically less relevant. The question was whether, at the high rotational speeds required to detect axial MRI, the fluid would not become turbulent even without a magnetic field. This argument was initially based on measurements taken by F. Wendt in 1933, but these had already been criticised as inadequate at the end of the 1950s. Using their own experimental setup in the form of a flat disc with split lids, researchers at Princeton had achieved high Reynolds numbers that led to very small increases in viscosity, many orders of magnitude below those theoretically expected for MRI.[16] His conclusion: "Good for MRI and dynamo."

After his conference contribution, Velikhov still had an ace up his sleeve, and what an ace it was. The newspaper La Sicilia had learned of his presence, contacted him and, on Tuesday, published almost an entire page with an interview of the "world-wide famous scientist, President of the Kurchatov Institute, advisor to the Russian President and member of the National Defence Council in Moscow," placed together with images of a nuclear submarine and a nuclear power plant. Defence Council of a leading nuclear goverment – none of the magnetohydrodynamicists present had ever sat at a table with such a mighty man – and I had ignored the last-minute invitation to the birthday party in Moscow because of a possibly closed cash register! As we now know, his political career began with Gorbachev when, as an ambitious party secretary, he asked to accompany him to London to meet Prime Minister Thatcher in 1984. She is said to have been happy with the two former fellow students from Lomonossov University – and Gorbachev had quickly found his future nuclear, defense and disarmament adviser with his rare combination of vision, diplomacy and expertise. Velikhov had just turned 50 when his old and/or new friend became the first man in the state the following year and sent him, Velikhov, to Chernobyl shortly afterwards as a test of his ability to limit the nuclear desaster.

---

[15] Princeton Plasma Physics Laboratory.
[16] H. Ji et al.: Nature 444 (2006), confirmed 2013 in BTU Cottbus in the lab of Egbers.

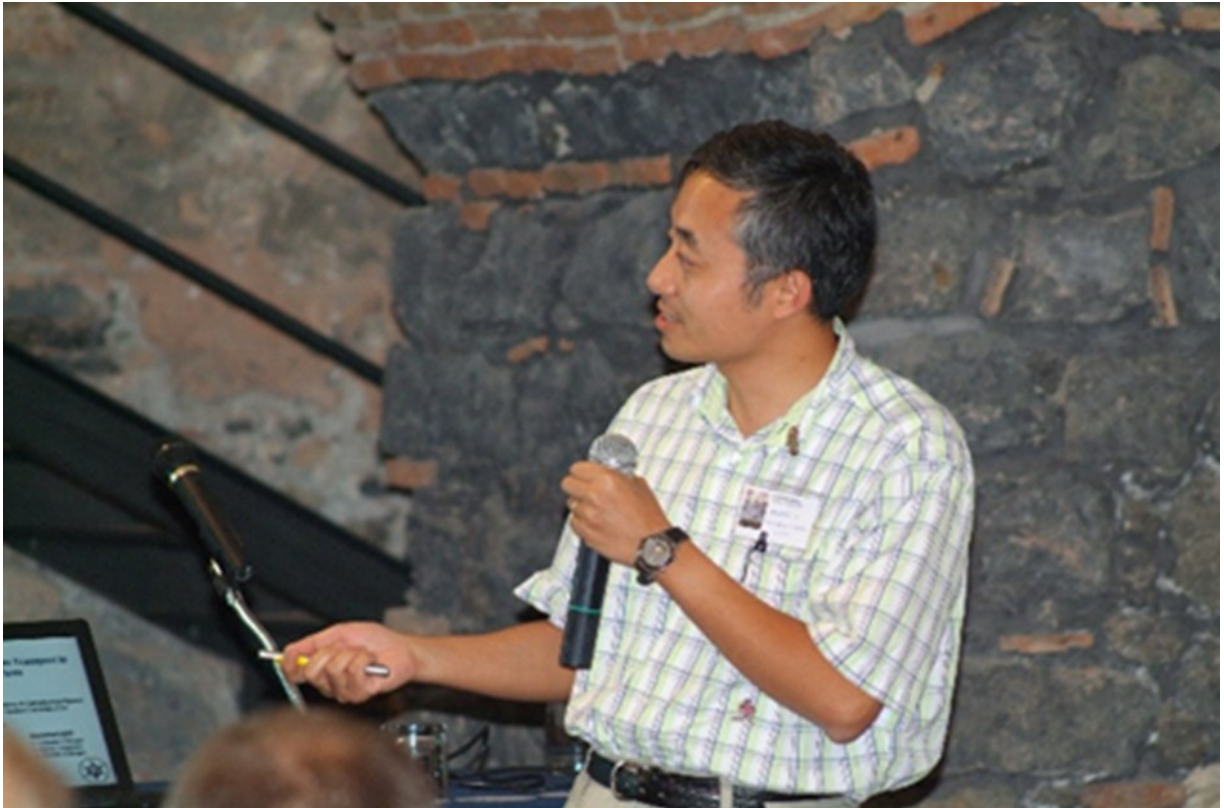

Figure 10 Catania 2007, Hantao Ji during his presentation

"Europe is not in danger of remaining cold and dark; the gas supply is guaranteed", the Velikhov first responded to the journalist's concerned question. "A second Chernobyl?

Impossible. There are already mini nuclear power plants that are safe and transportable and can supply electricity even to the most remote locations of the Earth. Nuclear fusion? It will be a reality in 10 years, although it will only become the planet's main energy source in 80 years. But the planet is already hungry for energy, with demand rising every day, especially in India and China. That is why power plants must be built on a massive scale to ensure sufficient electricity production. Russia has this experience, having built more than 200 nuclear submarines. This technology has resulted in a prototype power plant that can generate 70 megawatts, about the size of a boat. Russia will only lease the reactor, which will be under strict control. Our country, of course, can also supply the needed fuel. In the last five years, there have been only five minor accidents in Russia that were not related to the reactor or its core." Velikhov then takes the opportunity to present his life's work to the Italians and perhaps also to the conference participants: "In the case of nuclear fusion,[17] there is already a consortium of countries working to draw up an international protocol by the end of the year. The technology for building a fusion reactor is already complete. This includes the entire European Union and USA, China, Korea and Japan; it is a huge, expansive project. And we shall have realised Prometheus' great dream in a few decades. Realistically, it will take another 80 years before nuclear fusion can become the most important source of energy on our planet. For the rest of this century, we will therefore not be able to do without gas, oil, coal and nuclear power." In fact, Velikhov began a new career as Russia's representative on the multinational ITER Council[18] in November 2006 with the signing of the ITER agreement in Paris, chaired by President Jacques Chirac. Just a few days after our meeting in Catania, the giant project was officially launched on 24 October 2007 with the begin of the constructions.

---

[17] Velikhov had been leading research into controlled nuclear fusion in the USSR since 1973 (!).
[18] Chairman from 2009: Velikhov.

.

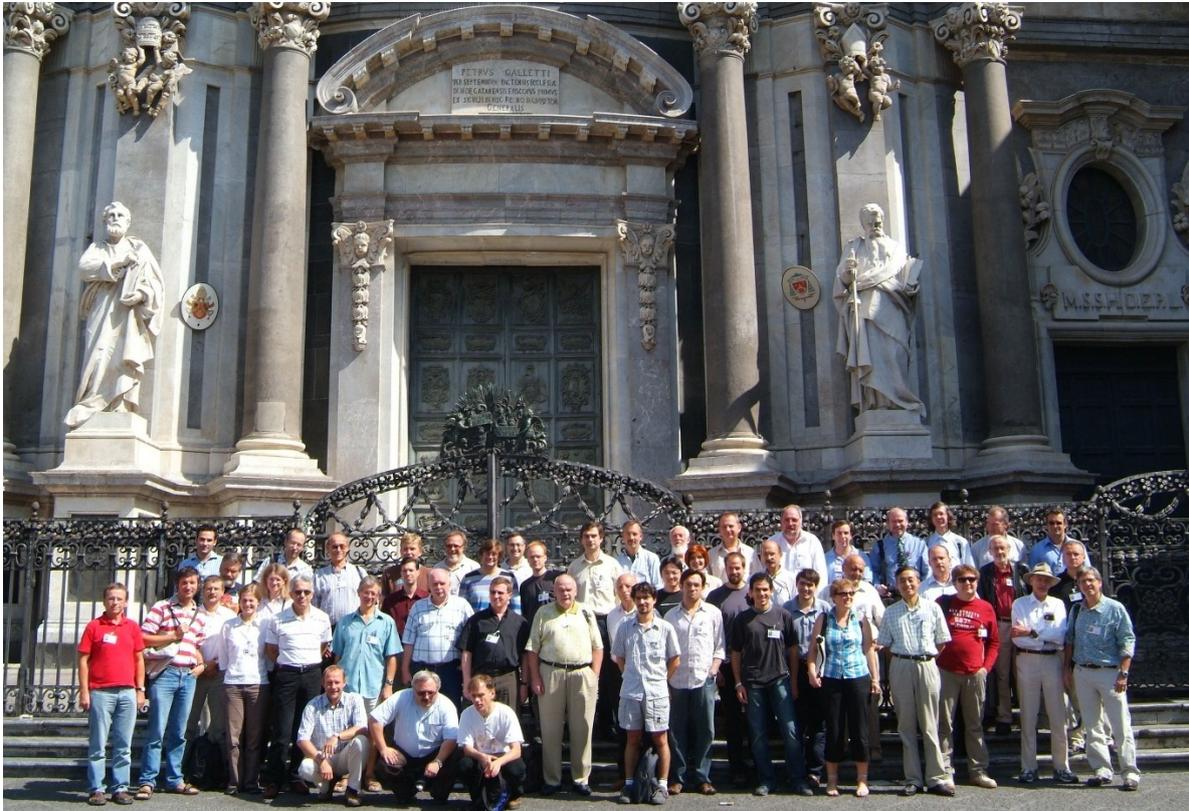

Figure 11 Conference photo 2007, Sant'Agata Cathedral, Piazza del Duomo, Catania

# 5. Epilog

Sometime in the 2010s, I received a message from Moscow that Velikhov was staying in Berlin and wanted to talk to me at his hotel. We had kept in loose contact over the years via the internet, and I was still eager to hear whether my negative calculations at the time had been correct or not. But nothing of the sort happened, at least not in direct speech. Over an elegant breakfast in his suite, he invited me to come to the Kurchatov Institute for a series of lectures on magnetohydrodynamics, because this field was being neglected in Russia, in both astrophysics and fusion research. He offered a considerable fee, but remained tight-lipped about accommodation and working conditions, possibly for security reasons. I agreed after he also accepted my suggestion to involve Rainer Hollerbach in the project, and we parted as allies. Later, he did not pursue the matter any further, and neither did I. Perhaps during the course of the conversation, it had become too ambitious or too inconvenient, or perhaps we, both pensioners, had overestimated our remaining suitability for high-flying plans. Then the Corona years came and even worse times. In December 2023, Ji, Stefani and me, three of the four participants in the MRI session at the 2007 MHD conference in Catania, met at the Babelsberg Observatory to exchange recent ideas. Velikhov, who said he believed "neither in God nor in Marx," had once again become unreachable during this time. Over the course of the year 2025, news came of his death in his nineties, significantly outside Russia while working at a Turkish nuclear power plant.

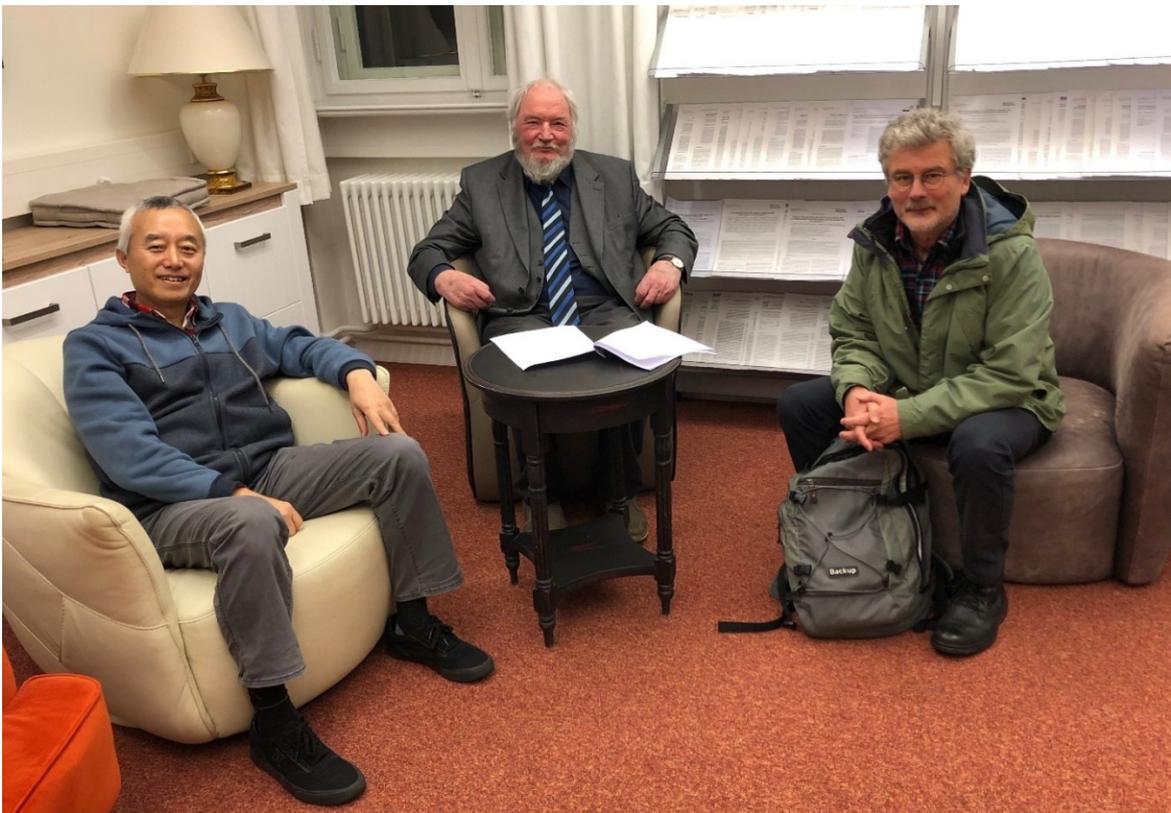

Figure 12 Hantao Ji (Princeton), Günther Rüdiger (Potsdam) and Frank Stefani (Dresden-Rossendorf) in a meeting room in the main building of the Babelsberg Observatory in December 2023

# 6. Acknowledgments

The author received material from this period from Alfio Bonanno, Hantao Ji and Frank Stefani, which they still found on their computers. He would also like to express his sincere thanks to the archivists at the newspaper La Sicilia for providing the interview with E.P. Velikhov from 2 October 2007. Almost all photos are from the archive of the author.